\documentstyle[prd,aps,psfig,subfigure,floats,preprint,tighten]{revtex}

\newcommand{\CL}{{\cal L}}

\newcommand{\bea}{\begin{eqnarray}}  \newcommand{\eea}{\end{eqnarray}}
\newcommand{\beq}{\begin{equation}}  \newcommand{\eeq}{\end{equation}}
\newcommand{\non}{\nonumber}  
\newcommand{\lmk}{\left(}  \newcommand{\rmk}{\right)}
  
\newcommand{\lhk}{\left \{ }  \newcommand{\rhk}{\right \} }
\newcommand{\del}{\partial}  

\newcommand{\bib}{\bibitem} \newcommand{\new}{\newblock}

\def\IB#1#2#3{{\bf #1}, #2 (19#3)}
\def\IBD#1#2#3{{\it ibid}. {\bf #1}, #2 (19#3)}
\def\NPB#1#2#3{Nucl. Phys. {\bf B#1}, #2 (19#3)}

\def\PLB#1#2#3{Phys. Lett. B~{\bf #1}, #2 (19#3)}
\def\PLBold#1#2#3{Phys. Lett. {\bf#1B}, #2 (19#3)}

\def\PRD#1#2#3{Phys. Rev. D~{\bf #1}, #2 (19#3)}

\def\PRL#1#2#3{Phys. Rev. Lett. {\bf#1}, #2 (19#3)}

\def\PTP#1#2#3{Prog. Theor. Phys. {\bf #1}, #2 (19#3)}

\def\JP#1#2#3{J. Phys. A~{\bf #1}, #2 (19#3)}

\begin{document}
\draft
\title{Evolution of a global string network in a matter-dominated universe}
\author{Masahide Yamaguchi}
\address{\it Department of Physics, School of Science,
    University of Tokyo, Tokyo~113-0033, Japan}
\author{Jun'ichi Yokoyama}
\address{\it Department of Earth and Space Science,
    Graduate School of Science, Osaka University, Toyonaka~560-0043,
    Japan} 
\author{M. Kawasaki}
\address{\it Research Center for the Early Universe, School of
    Science, University of Tokyo, Tokyo~113-0033, Japan}

\date{Oct 14 1999}
\maketitle
 
\begin{abstract}
  We evolve the network of global strings in the matter-dominated
  universe by means of numerical simulations. The existence of the
  scaling solution is confirmed as in the radiation-dominated universe
  but the scaling parameter $\xi$ takes a slightly smaller value,
  $\xi \simeq 0.6 \pm 0.1$, which is defined as $\xi = \rho_{s} t^{2}
  / \mu$ with $\rho_{s}$ the energy density of global strings and
  $\mu$ the string tension per unit length. The change of $\xi$ from
  the radiation to the matter-dominated universe is consistent with
  that obtained by Albrecht and Turok by use of the one-scale model.
  We also study the loop distribution function and find that it can be
  well fitted with that predicted by the one-scale model, where the
  number density $n_{l}(t)$ of the loop with the length $l$ is
  given by $n_{l}(t) = \nu/[t^2 (l + \kappa t)^2]$ with $\nu \sim
  0.040$ and $\kappa \sim 0.48$.  Thus, the evolution of the global
  string network in the matter-dominated universe can be well
  described by the one-scale model as in the radiation-dominated
  universe.
\end{abstract}



\label{sec:introduction}

Cosmic strings are formed in a class of cosmological phase transitions
\cite{KIB}.  Among them, gauged or local strings have been extensively
studied because they could produce density fluctuations responsible
for the large-scale structure formation and the cosmic microwave
background anisotropy \cite{VS}. A lot of analytical and numerical
studies have been done to confirm that after some relaxation period a
local string network enters the scaling regime. In this regime, the
large-scale behavior of a network scales with the Hubble radius and
the energy density of a network is given by
\beq
   \rho_{\rm string} = \xi \mu / t^2 \:,
\eeq
where $\xi$ is a constant irrespective of cosmic time.
\cite{AT,AT2,BB,AS}. Long strings intercommute to make loops
\cite{BB,AS,ACK}, which decay through radiating gravitational waves
\cite{VIL}. But an alternative scenario was recently proposed that
long strings directly emit massive particles and lose their own energy
\cite{VHS}. Thus, though the mechanism to drive a network to the
scaling regime is still in dispute, the existence of the scaling
solution is undoubted.

On the other hand, global strings have been less investigated and
considered only in the context of axion cosmology
\cite{GB,DA,flat,DQ,BS2,BS1}. It was assumed without direct
verification that a global string network relaxes into the scaling
solution just as the local counterpart, where long strings
intercommute to create loops with a typical size that is comparable
with the horizon scale and loops decay through emitting
Nambu-Goldstone~(NG) bosons.  However, since global strings have a
long-range interaction their dynamics could be different from that of
local strings.  In fact our numerical analysis of the evolution of a
complex scalar field in 2+1 dimensions revealed that the global
``strings'' do not relax into the scaling solution but the number of
defects per horizon volume increases logarithmically with time due to
the long-range interaction \cite{YYK}. In the previous paper
\cite{YKY} we examined the evolution of a global string network in the
radiation-dominated universe by numerically solving the equation of
motion for a complex scalar field representing a global string. Then
the above picture was confirmed. The scaling parameter $\xi$ takes a
constant value and becomes $\xi \sim (0.9-1.3)$, irrespective of the
cosmic time. Furthermore, one of the authors~(M.Y.) \cite{Yama} has
found that the number density $n_{l}(t)$ of the loop with the
length $l$ can be well fitted by the formula predicted from the
so-called one-scale model, that is, $n_{l}(t) = \nu/[t^{3/2} (l +
\kappa t)^{5/2}]$ with $\nu \sim 0.0865$ and $\kappa \sim 0.535$.
Thus, the evolution of a global string network in the radiation
dominated universe is well described by the one-scale model proposed
by Kibble \cite{KIB2,BEN,MAT}.

In this paper, for a complete understanding, we study the evolution of
a global string network in the matter-dominated universe, concretely,
whether it can be well described by the one-scale model as in the
radiation-dominated universe and, if at all, we clarify the relation
between scaling parameters in both eras.


We consider the following Lagrangian density for a complex scalar
field $\Phi(x)$,
\beq
  \CL[\Phi] = g_{\mu\nu}(\del^{\mu}\Phi)(\del^{\nu}\Phi)^{\dagger}
                 - V_{\rm eff}[\Phi,T] \:,
\eeq 
where $g_{\mu\nu}$ is identified with the Robertson-Walker metric and
the effective potential $V_{\rm eff}[\Phi,T]$ is given by
\beq
  V_{\rm eff}[\Phi,T] = \frac{1}{2}\lambda(\Phi\Phi^{\dagger} - \eta^2)^2 
                 + \frac{1}{3}\lambda T^2\,\Phi\Phi^{\dagger} \:,
  \label{eqn:effpot}
\eeq
which represents a typical second-order phase transition and the
$U(1)$ symmetry is broken below the critical temperature
$T_{c}=\sqrt{3}\eta$.

For cosmological purposes, it would be desirable to trace the
evolution of strings in the transition regime from the
radiation-dominated era to the matter-dominated era. Due to the
limitation of our computer powers, however, we concentrate on the
evolution of a global string network during the matter-domination
alone in this paper. Since the scaling property is expected to be
reached irrespective of initial conditions if at all, we start
simulations from a symmetric state with the equations of motion given
in the matter domination. Strings are formed soon and they evolve in
the matter-domination. As will be shown later, we confirm the scaling
behavior in the matter-dominated regime and find that $\xi$ in this
era is not different from that in the radiation-domination by more
than a factor of 2. Hence we expect that transition from the
radiation-dominated era to the matter-dominated era does not give rise
to any significant cosmological effects, justifying our approach.

In the matter-dominated universe, the equation of motion is given by
\beq
  \ddot{\Phi}(x) + 3H\dot{\Phi}(x) - \frac{1}{R(t)^2}\nabla^2\Phi(x)
   = - V'_{\rm eff}[\Phi,T] \:,
  \label{eqn:master}
\eeq
where the prime represents the derivative $\del/\del\Phi^{\dagger}$
and $R(t)$ is the scale factor which grows in proportion to $t^{2/3}$.
We define $\alpha(T)$ [$\alpha(T) > 1$] as $\alpha(T) \equiv \rho_{\rm
  mat}(T) / \rho_{\rm rad}(T)$ with $\rho_{\rm mat}(T)$ being the
contribution to the energy density from non-relativistic particles and
$\rho_{\rm rad}(T)$ being the contribution from relativistic particles
at the temperature $T$. Then,
\beq
  \alpha(T) = \alpha_c (T_{c}/T) \:,
\eeq
with $\alpha_{c} \equiv \rho_{\rm mat}(T_{c}) / \rho_{\rm rad}(T_{c})$.
Therefore, the Hubble parameter $H = \dot R(t)/R(t)$ and the cosmic
time $t$ are given by
\bea
  H^2 &=& \alpha(T) \frac{8\pi}{3 m_{\rm pl}^2} 
                     \frac{\pi^2}{30} g_{*} T^4 \\
      &=& \alpha_c \frac{4\sqrt{3}\pi^{3}\eta g_{*}}
                    {45 m_{\rm pl}^2} T^{3}, \\ 
   ~~~~~
  t &=& \frac{2}{3H} \equiv \frac{\epsilon}{T^{3/2}} \:,
  \label{eqn:hubble}
\eea
where $m_{\rm pl} = 1.2 \times 10^{19}$~GeV is the Plank mass and
$g_{*}$ is the total number of degrees of freedom for the relativistic
particles. We define the dimensionless parameter $\zeta$ as
\beq
  \zeta \equiv \frac{\epsilon}{\eta^{1/2}}  = \lmk \frac{5m_{\rm pl}^2}
      {\sqrt{3}\alpha_c \pi^3 g_{*} \eta^2}
  \rmk^{1/2} \:.
  \label{eqn:zeta}
\eeq
In our simulation, we take $\zeta = 8$ and $4$ to investigate $\zeta$
dependence on the result. We take the initial time $t_{i} =
t_{c}/(2\sqrt{2})$ corresponding to $T_{i}=2T_{c}$ and the final time
$t_{f} = 70(140)\,t_{i}$, where $t_c$ is the epoch with $T=T_c$. Since
the $U(1)$ symmetry is restored at the initial time $t = t_{i}$, we
adopt as the initial condition the thermal equilibrium state with the
mass squared,
\beq
  m^2 = \left. \frac{d^2 V_{\rm eff}[|\Phi|,T]}{d|\Phi|^2} 
        \right|_{|\Phi|=0} \:,
\eeq
which is the inverse curvature squared of the potential at the origin
at $t = t_{i}$.

Below we measure all of the physical quantities in units of $t_{i}$.
Then the equation of motion is given by
\beq 
\ddot{\Phi}(x) + \frac{2}{t}\dot{\Phi}(x) -
  \frac{1}{t^{4/3}}\nabla^2\Phi(x) 
   = - \lmk |\Phi|^2 + \frac{\zeta^2}{6\sqrt{3}\,t^{4/3}} 
                                - \frac{\zeta^2}{24\sqrt{3}} \rmk 
    \Phi^{\dagger} \:, 
\eeq
where $\lambda$ is set to unity for brevity. The scale factor $R(t)$
is normalized as $R(1) = 1$.

Using the second-order leap-frog method (see Ref. \cite{Yama} for
details), we evolve the global string networks in the matter-dominated
universe. In order to judge whether the global string network relaxes
into the scaling regime in the matter-dominated universe, we give time
development of the scaling parameter $\xi$ defined as $\rho = \xi \mu
/ t^2$. In our simulations, a lattice is identified with a part of a
string core if the potential energy density there is larger than that
corresponding to the field value of a static cylindrically-symmetric
solution at $r = \delta x_{\rm{phys}}/\sqrt{2}$.

We perform the simulations in seven different sets of lattice sizes,
spacings, and $\zeta$~(see Table \ref{tab:set1}, \ref{tab:set2}). In
cases (1) and (5), the box size is nearly equal to the horizon volume
$(H^{-1})^{3}$ and the lattice spacing to a typical width $\delta \sim
1.0/(\sqrt{2}\eta)$ of a string at the final time $t_{f}$. For each
case, we simulate the system from 10~(Eqs. (1)-(6)) or 300~(Eq. (7))
different thermal initial conditions.

If the simulation box is much larger than the horizon volume, it is
reasonable to think that the boundary effect is negligible. But, in
our simulation, the simulation box is comparable with or at most
$4^{3}$ times as large as the horizon volume at the final time of the
simulation so that we should be careful to avoid possible boundary
effects. In fact, a long-range force works between global strings so
that the boundary effect cannot necessarily be neglected if the number
of long strings in the simulation box is very small. As shown later,
there are only a few long strings in our case so that the boundary
effect could be significant.  Hence in this paper we run simulations
with two different boundary conditions of distinct features and employ
a large enough simulation box so that the results with the different
boundary conditions converge to each other. We thereby obtain a result
free from the boundary effects.

\begin{figure}
  \begin{center}
    \leavevmode\psfig{figure=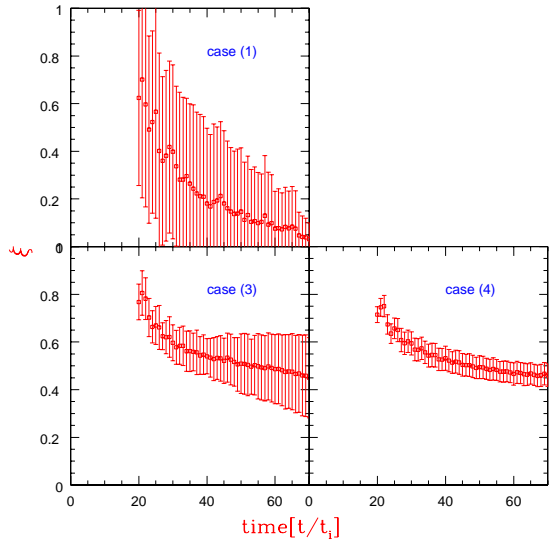,width=8cm}
  \end{center}
  \caption{Time development of $\xi$ under the periodic boundary
    condition. Symbols~($\Box$) represent time development of $\xi$.
    The vertical lines denote a standard deviation.}
  \label{fig:periodic}
\end{figure}

First, we adopt the periodic boundary condition, under which a string
feels an attractive force from the boundary and there exists no
infinite string so that strings can completely disappear in the
simulation box if the Hubble radius becomes larger than the dimension
of the simulation box.  Figure \ref{fig:periodic} represents time
development of $\xi$ with $\zeta = 8$~(cases(1),(3),(4)) under this
boundary condition. In case (1) with the smallest box, the boundary
effect is so significant that strings tend to disappear, which is
inconsistent with the result in Refs. \cite{KIB2,BEN,MAT}. The larger
the box size is, the less important the boundary effect is. In case
(3) corresponding to the largest box simulations, the boundary effect
is less significant so that $\xi$ seems to relax to a constant with
$\xi \sim 0.49 \pm 0.02$.

\begin{table}
\caption{Three different sets of the simulations under the periodic
  boundary condition.}
\label{tab:set1}
  \begin{center}
     \begin{tabular}{ccccccc}
         Case & Lattice &
         Lattice spacing & $\zeta$ & Realization & ${\rm Box~size}/H^{-1}$
         & $\xi$ \\
         & number & [unit = $t_{i}R(t)$] &  &  & (at final time) &  \\
        \hline
        (1) & $128^3$ & $\sqrt[3]{70}/85$ & 8 & 10 & 1(at 70) &
        Disappearance \\ 
        (3) & $256^3$ & $\sqrt[3]{70}/85$ & 8 & 10 & 2(at 70) &
        $0.50\pm0.02$ \\
        (4) & $256^3$ & $2\sqrt[3]{70}/85$ & 8 & 10 & 4(at 70) &
        $0.49\pm0.02$ \\
     \end{tabular}
  \end{center}
\end{table}
\begin{table}
\caption{Seven different sets of the simulations under the reflective
  boundary condition.}
\label{tab:set2}
  \begin{center}
     \begin{tabular}{ccccccc}
         Case & Lattice &
         Lattice spacing & $\zeta$ & Realization & ${\rm Box~size}/H^{-1}$
         & $\xi$ \\
         & number & [unit = $t_{i}R(t)$] &  &  & (at final time) &  \\
        \hline
        (1) & $128^3$ & $\sqrt[3]{70}/85$ & 8 & 10 & 1(at 70) &
        $1.93\pm0.04$ \\ 
        (2) & $128^3$ & $2\sqrt[3]{70}/85$ & 8 & 10 & 2(at 70) &
        $1.03\pm0.03$ \\
        (3) & $256^3$ & $\sqrt[3]{70}/85$ & 8 & 10 & 2(at 70) &
        $1.13\pm0.02$ \\
        (4) & $256^3$ & $2\sqrt[3]{70}/85$ & 8 & 10 & 4(at 70) &
        $0.72\pm0.03$ \\
        (5) & $128^3$ & $\sqrt[3]{140}/85$ & 4 & 10 & 1(at 140) &
        $1.36\pm0.06$ \\
        (6) & $256^3$ & $\sqrt[3]{140}/85$ & 4 & 10 & 2(at 140) &
        $1.21\pm0.02$ \\
        (7) & $128^3$ & $2\sqrt[3]{70}/85$ & 4 & 300 & 2(at 70) &
        $1.25\pm0.02$ \\
     \end{tabular}
  \end{center}
\end{table}

\begin{figure}
  \begin{center}
    \leavevmode\psfig{figure=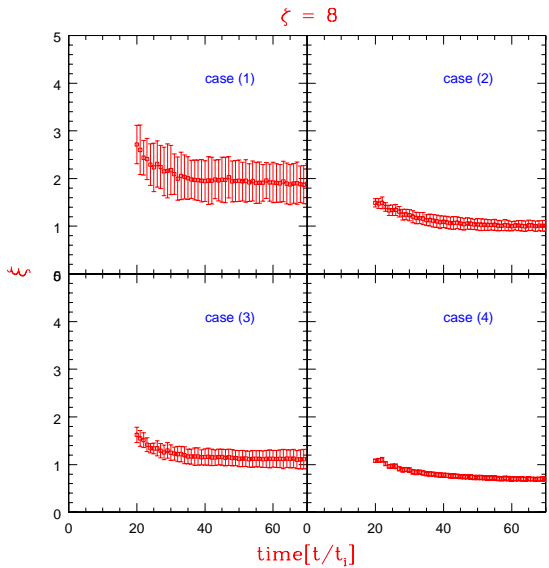,width=8cm}
  \end{center}
  \caption{Time development of $\xi$ in the cases from (1) to
    (4) with $\zeta=8$ under the reflective boundary condition.
    Symbols~($\Box$) represent time development of $\xi$.  The vertical
    lines denote a standard deviation over different initial
    conditions.}
  \label{fig:reflective}
\end{figure}

\begin{figure}
  \begin{center}
    \leavevmode\psfig{figure=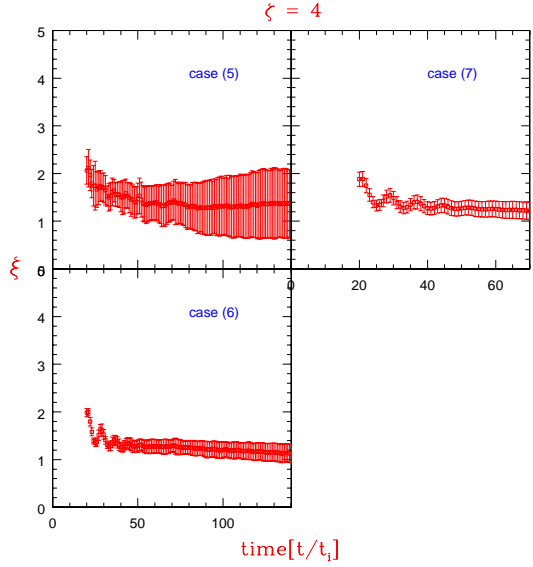,width=8cm}
  \end{center}
  \caption{Time development of $\xi$ in the cases from (5) to
    (7) with $\zeta=4$ under the reflective boundary condition.}
  \label{fig:reflective2}
\end{figure}

Next we adopt the reflective boundary condition, where
$\nabla^{2}\Phi(x)$ on the boundary points disappears.  Under this
boundary condition, a string suffers a repulsive force from the
boundary so that a string near the boundary intercommutes less often
than that near the center of the simulation box because the partner to
intercommute only lies in the inner direction of the boundary. Thus,
the number of the strings tends to be more than that in the real
universe. Figures \ref{fig:reflective} and \ref{fig:reflective2}
represent time development of $\xi$ with $\zeta = 8$~[cases(1)-(4)]
and $\zeta = 4$~[cases(5)-(7)], where $\xi$ becomes a constant
irrespective of time with (1)~$1.93\pm0.04$, (2)~$1.03\pm0.03$,
(3)~$1.13\pm0.02$, (4)~$0.72\pm0.03$, (5)~$1.36\pm0.06$,
(6)~$1.21\pm0.04$, and (7)~$0.72\pm0.03$.  $\xi$ has larger values in
smaller-box simulations due to the boundary effect as explained above.
Also, for larger-box simulations, $\xi$ depends very little on
$\zeta$.

From the results of the largest-box simulations containing the largest
number of the Hubble volume at the final epoch, we conclude that $\xi$
converges to a constant $\xi \simeq 0.6 \pm 0.1$ irrespective of the
boundary conditions.  This result will be supported later by comparing
it with our previous results in the radiation-dominated regime
\cite{YKY,Yama} through an analytic model of Albrecht and Turok
\cite{AT}.


Next we investigate the loop distribution, which is predicted by
Kibble's one-scale model \cite{KIB2,BEN,MAT} as
\beq 
  n_{l}(t) = \nu / 
     [t^2 (l + \kappa t)^2] \:,
\eeq
where $\nu$ is a constant, $l$ is the length of a loop, and the
logarithmic dependence of $\mu$ is neglected.  In contrast with local
strings, the dominant energy loss mechanism of global strings is the
radiation of the associated Nambu-Goldstone field \cite{GB}. We define
the radiation power $P$ as $P = \kappa \mu$ where $\kappa$ is a
constant.

\begin{figure}
  \begin{center}
    \leavevmode\psfig{figure=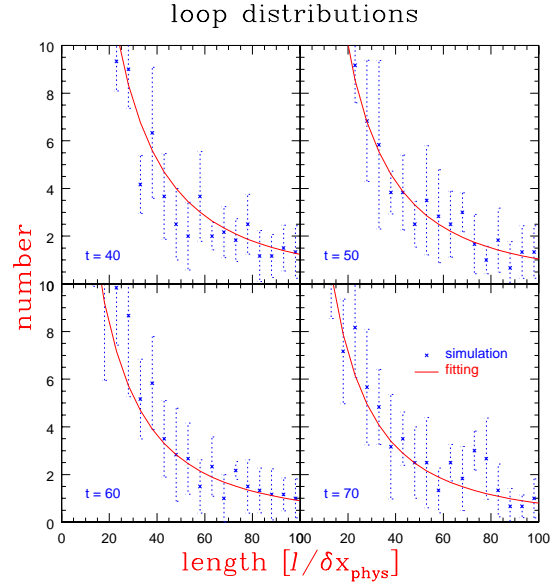,width=8cm}
  \end{center}
  \caption{Loop distributions at $t = 40, 50, 60, 70$ are
    depicted. The number is summed over the box size [$128(\delta
    x)^3$] and 50 realizations for each groups. Bins are cut every
    5$\times\delta x$.}
  \label{fig:loop}
\end{figure}

\begin{figure}
  \begin{center}
    \leavevmode\psfig{figure=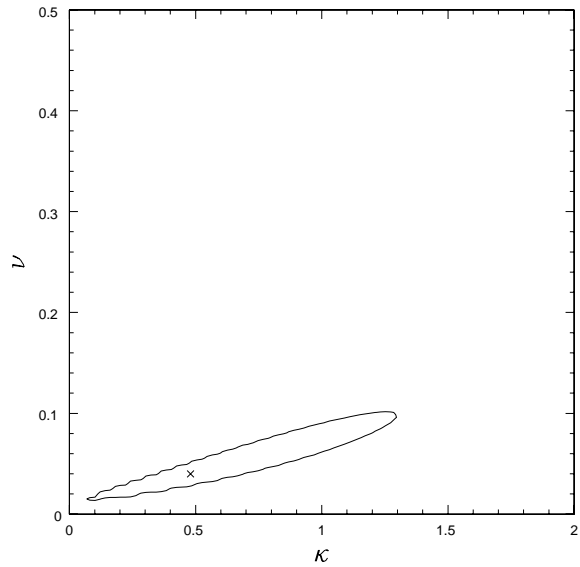,width=8cm}
  \end{center}
  \caption{Fittings for $\kappa$ and $\nu$. The cross point represents the
    best fit values for $\kappa$ and $\nu$. The solid circle denotes
    $68\%$ C.L.}
  \label{fig:kappanu}
\end{figure}

We determine whether the loop distribution in the simulation coincides
with the above function. Since case (4) with the largest box takes too
much time for one realization, we investigate the loop distribution
for case (7) under the reflective boundary condition.\footnote{$\xi$
  in case (7) is about twice as large as that in case (4)
  corresponding to the largest box simulation so that $\nu$ in the
  real world may become half as large as that obtained in our
  simulation.}  The loop distribution is depicted in Fig.\ 
\ref{fig:loop} in case (7) at $t = 40, 50, 60,$ and 70. Since long
strings are rare, we cut the length of loops into bins with the width
5$\times\delta x$. Then, we divide 300 realizations into 6 groups
comprised of 50 realizations and we summed the number of loops over 50
realizations for each groups. The dot represents the number of loops
averaged over six groups and the dashed line represents the standard
deviation. They can be simultaneously fitted with the above formula if
one takes $\nu \sim 0.040$ and $\kappa \sim 0.48$. Fittings for
$\kappa$ and $\nu$ are also given in Fig. \ref{fig:kappanu}. Thus, the
loop production function as well as the large scale behavior of the
string scales together for the global string network. Note that
$\kappa$ takes almost the same value both in the matter- and
radiation-dominated universe, which implies that NG bosons emission is
not affected by the background universe and the large-scale behavior
of a string network but is decided by the fundamental physics near a
string segment.

\indent

As discussed in \cite{AT}, the one-scale model can predict the scaling
parameter $\xi_{m}$ in the matter domination from $\xi_{r}$ in the
radiation domination. From the Nambu-Goto action together with the
intercommutation effect, the evolution of the string energy density
$\rho$ in the expanding universe is given by
\bea
  d\rho/dt &=& -3H\rho + (1-2V^{2})H\rho - c\rho/L,
  \non \\
                   &=& -3H\rho + H^{2} L \rho - c\rho/L,
\eea
where $L = t / \sqrt{\xi}$ is the characteristic scale with $\rho =
\mu/L^{2}$, $V^{2}$ is the average velocity squared of a string, and
$c$ is a constant representing the intercommutation
efficiency.\footnote{Though $c$ may be dependent on $V$, we set $c$ to
  be a constant as the zeroth order approximation.} Defining $\gamma$
as $\gamma = H^{-1}/L$, the evolution of $\gamma$ is given by
\beq
  \frac{d\gamma}{dt} = -\frac{H}{2} \lhk c\gamma^{2} - 
                         [2\,\dot{(H^{-1})} - 3]\gamma - 1 \rhk.
\eeq
Then, in the radiation-domination with $\dot{(H^{-1})}=2$, we obtain
the fixed point $\gamma_{r}$ given by
\beq
  \gamma_{r} = (1+\sqrt{1+4c})/2c.
\eeq
On the other hand, in the matter domination with $\dot{(H^{-1})}=3/2$,
we obtain the fixed point $\gamma_{m}$ given by
\beq
  \gamma_{m} = 1/\sqrt{c}.
\eeq
Considering $\xi_{r}=\gamma_{r}^{2}/4$ and
$\xi_{m}=4\gamma_{m}^{2}/9$, we find the relation between $\xi_{r}$
and $\xi_{m}$,
\beq
  \xi_{m} = \frac{16}{9} \frac{\xi_{r}}{(2\sqrt{\xi_{r}}+1)}.
\eeq
Putting $\xi_{r} \sim (0.9-1.3)$ \cite{Yama} into the relation,
$\xi_{m}$ is predicted to be $0.6-0.7$, which coincides with our
numerical results.

In this paper, we investigated the evolution of a global string
network in the matter-dominated universe. The network relaxes into the
scaling regime as in the radiation-dominated universe but the scaling
parameter $\xi_{m}$ takes a smaller value, $\xi_{m} \simeq 0.6 \pm 0.1$,
which is consistent with the value predicted from $\xi_{r}$ by use of
the formula obtained by Albrecht and Turok \cite{AT}. The loop
distribution is also obtained and compared with that predicted by the
one-scale model, where the number density $n_{l}(t)$ of the loop
with the length $l$ is given by $n_{l}(t) = \nu/(t^2 [l + \kappa
t)^2]$. With $\nu \simeq 0.040$ and $\kappa \simeq 0.48$, the loop
distribution function can be well fitted with that predicted by the
one-scale model. Thus, the evolution of the global string network in
the matter-dominated universe can be well described by the one-scale
model as in the radiation-dominated universe.

This work was partially supported by the Japanese Grant-in-Aid for
Scientific Research from the Monbusho, Nos.\ 10-04558 (M.Y.),
11740146 (J.Y.), and ``Priority Area: Supersymmetry and Unified Theory of
Elementary Particles (\# 707)'' (J.Y. and M.K.).

\end{document}